\def\3he{$^3$He}
\def\4he{$^4$He}
\def\7li{$^7$Li}
\def\6li{$^6$Li}
\def\Teff{$\rm T_{eff}$}

\documentclass{iaus}
\usepackage{graphicx}

\title[Li  isotopes in metal-poor halo dwarfs] %% give here short title %%
{Li  isotopes in metal-poor halo dwarfs,\\ a more and more complicated 
story}

\author[Monique Spite \&  Fran\c cois Spite]   %% give here short author list %%
{Monique Spite \and  Fran\c cois Spite
}

\affiliation{ GEPI, Observatoire de Paris, 92195 Meudon Cedex, CNRS UMR 8111 \\ 
email: {\tt monique.spite@obspm.fr, francois.spite@obspm.fr }}

\pubyear{2010}
\volume{268}  %% insert here IAU Symposium No.
\jname{Light elements in the Universe}
\editors{C. Charbonnel, M. Tosi, F. Primas \& C. Chiappini, eds.}

\begin{document}

\maketitle

\begin{abstract}
The nuclei of the lithium isotopes  are fragile, easily destroyed, so that,  at variance with most of the other elements, they cannot be formed in stars through steady hydrostatic nucleosynthesis.

The \7li isotope is synthesized during primordial nucleosynthesis in the first minutes after the Big Bang  and later by cosmic rays, by novae and in pulsations of AGB stars (possibly also by the  $\nu$ process).
\6li is mainly formed by cosmic rays. The  oldest (most metal-deficient) warm galactic stars should retain the signature of these processes if, (as it had been often expected) lithium is not depleted in these stars. 
The existence of a "plateau'' of the  abundance of \7li (and of its slope) in the warm metal-poor stars is discussed.  At very low metallicity ($\rm [Fe/H]<-2.7 dex)$ the star to star scatter increases significantly towards low Li abundances.  
The highest value of the lithium abundance in the early stellar matter of  the Galaxy ($\rm log \epsilon(Li)$ = A(\7li) = 2.2 dex\footnote{In the literature the lithium abundance, is noted indifferently by $\rm log \epsilon(Li)$ , or by A(Li); both notations are in logarithmic scale of number of atoms where $\rm log \epsilon(H)= A(H) =12$}
) is much lower than the the value  ($\rm log \epsilon(Li)$ = 2.72) predicted by the standard Big Bang nucleosynthesis,  according to  the specifications found by the satellite  WMAP.  After gathering a homogeneous stellar sample, and analysing its behaviour, possible explanations of the disagreement between Big Bang and stellar abundances are discussed (including early astration and  diffusion). On the other hand, possibilities of lower productions of  \7li  in the standard and/or non-standard Big Bang nucleosyntheses are briefly evoked.

 A surprisingly high value (A(\6li)=0.8 dex) of the abundance of the \6li isotope has been found in a few warm metal-poor stars. Such a  high abundance of \6li independent of the mean metallicity in the early Galaxy cannot be easily explained. But are we really observing \6li~?

\keywords{lithium, stellar abundances, primordial nucleosynthesis, Pop II stars.}
\end{abstract}

\firstsection % if your document starts with a section,

%%%%% Section 1
\section{Introduction}
%**************
\subsection{A first view} 

In the eighties and the nineties,  the abundance of  \7li in the warm ($\rm 5700<Teff<6500K$) metal-poor dwarfs was found  constant, independent of the temperature and of the metallicity, at least in the interval $\rm -2.8<[Fe/H]<-1.8$ : Spite \& Spite (\cite{SPI82Nat}, and \cite{SPI82AA}) and subsequent papers (e. g. Spite, Spite and Maillard  (\cite{SSM84AA}), Hobbs \& Thorburn (\cite{HoT91AP}), Molaro,  Primas  and  Bonifacio (\cite{MPB95AA}), Spite et al. ( \cite{SFN96AA}),
 Bonifacio and Molaro (\cite{BoM97AA}), Smith et al. (\cite{SLN98AA})). Later works,  published up to 2005,  are vividly reviewed in the section 2 of Charbonnel and Primas (\cite{CP2005}) : the small differences between authors are discussed.
 
This "universal" behaviour of lithium suggested that \7li observed in the old metal poor dwarfs was synthesized during
primordial nucleosynthesis (when the Universe was only a few minutes 
old) and that  lithium, although a very fragile element, had survived unaltered in the atmosphere of warm metal-poor dwarfs.  In the so called "standard" stellar models (i. e. without diffusion and mixing), the Li depletion is negligible (Deliyannis et al. \cite{DDK90}, Pinsonneault et al. \cite{PDDAS}).
However, already at that time, several theoreticians of stellar atmospheres thought that it was difficult to admit that the abundance of lithium in the atmosphere of the old very metal-poor dwarfs had remained unchanged during 13 billions years.
Reciprocally, it was also difficult to explain a strictly uniform lithium depletion in the warm metal-poor dwarfs, whatever their temperature, mass and metallicity.

The primordial abundance of \7li had been  then largely identified with the stellar abundance found on the plateau,  the value being between $\rm log \epsilon(Li) = 2.0$  and $\rm log \epsilon(Li) = 2.3$
depending on the temperature scale adopted by the authors. Moreover, taking into account the errors of measurement and the uncertainties of the observed abundances, the simultaneous production of all the light elements (\4he,  \3he, D and Li) by the Big Bang, could be explained.\\

In spite of the obvious difficulty of disentangling the  \6li and \7li profiles in stellar spectra, attempts were made at finding the abundance of  \6li  : it was possible only in a handful of metal-poor dwarfs. 
For the first time Smith Lambert \& Nissen (\cite{SLN93AA}), claimed a positive detection of  \6li in a metal-poor star: HD84937 ($\rm[Fe/H]\approx-2.1$).
Later Nissen et al., (\cite{NLP99AA}) observed \6li  in 2 metal-poor stars  out of five; see also Cayrel et al. (\cite{Cay99}). The existence of \6li  (more fragile that \7li) in some warm metal-poor dwarfs reinforced the idea that \7li is not depleted in such stars.

%**************
\subsection{A second step} 
Two astronomical satellites brought  new  points of view :\\ 
\noindent- in 1997, Hipparcos has provided accurate parallaxes for numerous stars, improving the determination of the parameters of the stellar atmospheres (and enabling to discriminate dwarfs and subgiants).\\
- in 2003, WMAP has determined (Benett et al. \cite{BCL03}) the  physical conditions of the Big Bang especially a precise value for the baryons to photons ratio : $\eta$, leading to precise values of the abundances of the light elements produced in the standard Big Bang, including lithium. This Li abundance  was found much higher than the abundance measured in the warm metal-poor dwarfs and thus the interest for a precise re-determination of the lithium abundance was reactivated.

%%%%  Section 2
\section{The lithium isotopes of in the matter of the early Galaxy}

%**************
\subsection{Formation of Lithium}  
A peculiarity of  lithium is that it is very fragile and destroyed at "low" temperatures.  Unlike most of the other elements, when it is synthesized inside the stars by hydrostatic nucleosynthesis, it is immediately destroyed.
Lithium is formed :\\ 
-- during the Big Bang (primordial nucleosynthesis) (\7li and negligible amount of \6li).\\ 
-- by spallation (cosmic rays  or in superbubbles: mainly \6li)\\
-- possibly by the ``$\nu$'' process in type II supernovae  (\7li)\\

\noindent Pulsations of AGB stars and novae explosions can also enrich the interstellar medium in \7li, but  not  in the early times  of the Galaxy.

%**************
\subsection{Primordial abundance of  \7li}  
The  \7li observed in the atmospheres of the very old galactic stars has been mainly formed by the primordial nucleosynthesis. Cosmic rays and $\nu$ process in massive supernovae could produce a small increase of $\rm \epsilon(^7Li)$ with metallicity (a slope).

In the standard Big-Bang, the production of lithium depends only on the  baryons to photons ratio $\eta$. This ratio has been deduced, with precision, from the measurements of the cosmic microwave background radiation by the satellite WMAP  :  Benett et al. (\cite{BCL03}),  Spergel et al. (\cite{SBD07}), Cyburt et al. (\cite{CFO08}) :  $\eta = 6.23  \pm 0.17~  10^{-10}$.
 
Then, the primordial lithium abundance is $\rm log \epsilon(^7Li) = 2.72 \pm 0.06$  for  $\rm log \epsilon(H)=12$ (see also Komatsu et al., \cite{KDN09AS}, and Iocco et al., \cite{IMM09},  for the discussion of the uncertainties).

\subsection{Depletion of lithium in the atmosphere of the stars}
It is generally admitted that the abundances measured in the atmospheres of unevolved stars are representative of  the abundances in the material from which the star has been originally formed.\\
However different phenomena are known to affect the superficial abundance of lithium during the life of the star.

Lithium is a very fragile element destroyed as soon as the temperature is higher than $\rm 2.5 ~10^6 K $  for \7li and even  $\rm 2.0 ~10^6 K $ for \6li. 

If there is some mixing between the surface of the star and the hot deep layers, little by little lithium is destroyed and disappears from the atmosphere of the star.
In giants, Li is strongly diluted after the first dredge-up.
Even in dwarfs lithium is often depleted: in the atmosphere of the Sun after 4.5 Gyr, the lithium abundance is only $\rm log \epsilon(Li) = 1.03$ (Caffau et al., \cite{Caf09}) although in meteorites $\rm log \epsilon(Li)$ reaches $\approx$ 3.25 (a value representative of the abundance of lithium in the material that formed the Sun). 
However  in the warm metal-poor dwarfs/turnoff stars, mixing is not as deep as in solar metallicity stars, and lithium could have been preserved.

Lithium is also supposed to slowly settle down into the stars by diffusion (gravitational settling), but the diffusion can be easily thwarted by turbulent mixing. 

%%%%  Section 3
\section{The behaviour of \7li in the most metal poor stars in the Milky Way}

The difference between the lithium abundance  predicted by BBN + WMAP and the mean value observed in the atmosphere of the old galactic dwarfs, lead to a redetermination of the lithium abundance, with an effort towards better temperature determinations: the determination of  lithium abundance is very sensitive to the choice of the temperature of the stellar atmosphere. 
Moreover, the precise behaviour of $\rm log \epsilon(Li)$  vs. temperature and metallicity and also the scatter around the mean relations may give a clue about the exact mechanism of lithium depletion.

Since 2005  several papers about the lithium abundance in very metal-poor stars were published. They all try to determine very carefully the temperatures of the stellar atmospheres, using high quality data: it is essential to discriminate between intrinsic scatter and determinations errors.

\noindent--Charbonel \& Primas (\cite{CP2005}) have determined the temperatures from a very precise photometry ({\it uvby}) associated with the Alonso  et al. (\cite{AAM96}) calibration and taking into account very carefully the reddening of the stars. 
%They used the Hipparcos data in order to  discriminate dwarfs from subgiants, finding  that the subgiants (near the turnoff) had a slightly larger Li abundance.
Mel\'endez  et al. (\cite{MCR09a}, \cite{MCR09b}) had a similar approach but they used a new implementation of the infrared flux method to determine the temperatures from multiband IR photometry.\\
--Boesgaard et al. (\cite{BSD05}) and Hosford et al. (\cite{HRG09}) based their analysis on excitation temperature (the abundances from the Fe I lines must be independent of the excitation potential of the line).\\
--All the other authors used the wings of the hydrogen lines (mainly H$\alpha$) to determine the temperatures: Asplund et al. (\cite{ALN06}), Bonifacio et al. (\cite{BMS07}), Gonz\'alez Hern\'andez et al. (\cite{GBL08}), Garc\'{i}a P\'erez et al. (\cite{GCR08}, \cite{GAI09}), Aoki et al.( {\cite{ABB09}).\\

In Fig. \ref{Figlit}  we present  $\rm log \epsilon(Li)$ versus \Teff,  for the metal-poor dwarfs and turnoff stars  with $\rm [Fe/H]< -2.0$, measured in these different works.  There is a rather good agreement between the different determinations, (even if some slight systematic differences appear between different authors). No significant trend of $\rm log \epsilon(Li)$ vs. \Teff~is apparent. The mean value of the lithium abundance is $\rm log \epsilon(Li) \approx 2.15$ : about  four times less than the abundance  formed by the standard Big Bang nucleosynthesis.

This mean value of the lithium abundance measured in the old galactic stars is also very close to the value  measured in $\rm \omega~Cen$  (see Bonifacio et al., \cite{BMSs268}, in this symposium): $\rm log \epsilon(Li) \approx 2.19 \pm 0.14$ . The cluster $\rm \omega~Cen$ is now considered as the remnant of a captured dwarf galaxy. 
As a consequence  $\rm log \epsilon(Li)=2.15 $ could be a "universal'' value of stellar Li abundance at low metallicity.

%FIG 1
\begin{figure}[t]
\begin{center}
\resizebox{12.0cm}{6.5cm}
{\includegraphics{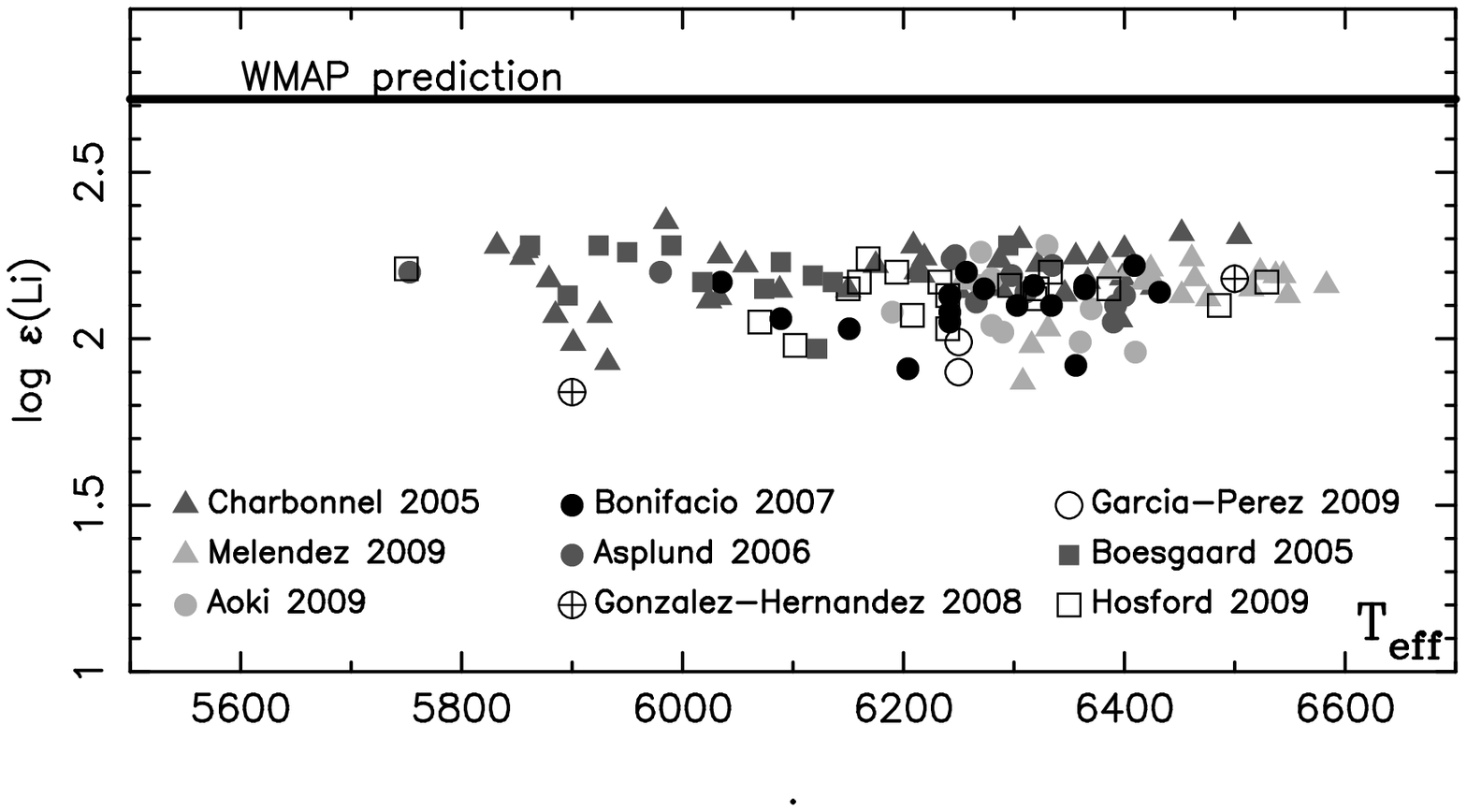}}
\resizebox{11,5cm}{3.0cm}
{\includegraphics{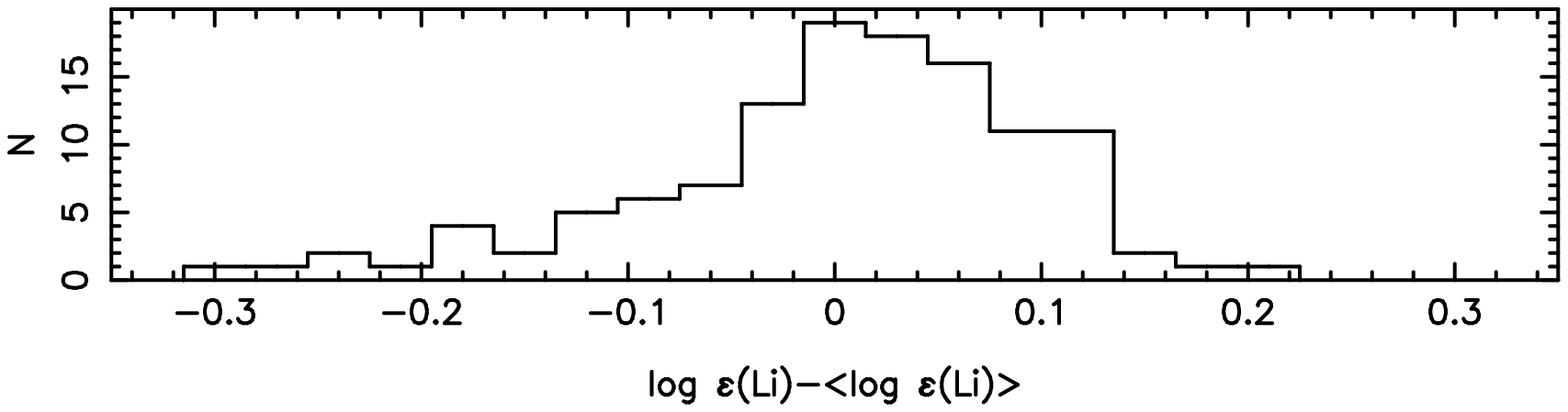}}
 \caption{Lithium abundance versus Temperature for metal-poor stars
 ($\rm[Fe/H]<-2$) and (below) histogram. Triangles indicate that the temperature has been
 determined from photometry, circles from the profile of the hydrogen
 lines and squares from the excitation temperature. In this interval
 of temperature and metallicity, the lithium abundance is independent of the temperature but the scatter is rather large. The mean value of the lithium abundance is 
 $\rm log \epsilon(Li) \approx 2.15$, more than 0.55 dex below the prediction of  standard BB +  WMAP (full black line).
On the histogram showing the distribution of the distances of each point to the mean value of the lithium abundance it can be seen that the distribution is not gaussian.}
\label{Figlit}
\end{center}
\end{figure}

%FIG 2
\begin{figure}[h]
\begin{center}
\resizebox{12.0cm}{6.5cm}
{\includegraphics{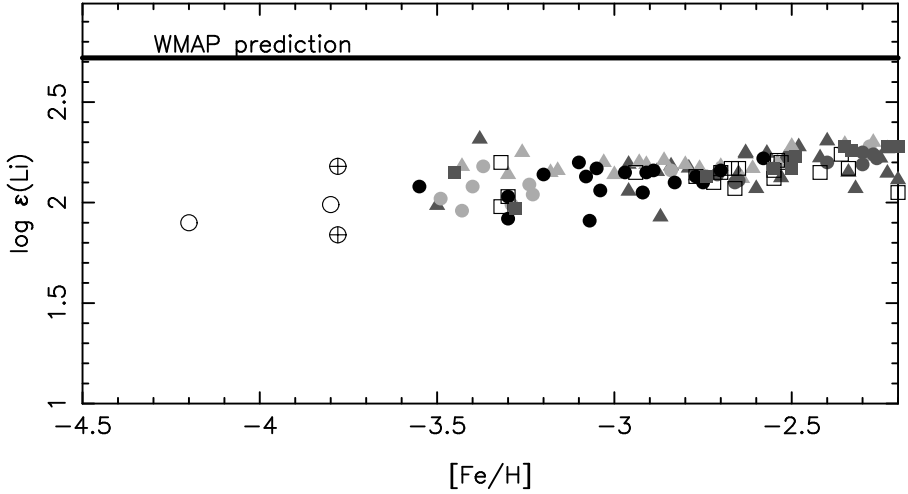}}
 \caption{$\rm log \epsilon(Li)$ vs. [Fe/H], for turnoff stars (\Teff $>5800$K). The symbols are the same as in Fig. 1. The slight decline of  $\rm log \epsilon(Li)$ when the metallicity decreases 
 appears for $\rm [Fe/H] < -3$.
}
\label{lilogfe}
\end{center}
\end{figure}

%FIG 3
\begin{figure}[h]
\begin{center}
\resizebox{12.0cm}{6.5cm}
{\includegraphics{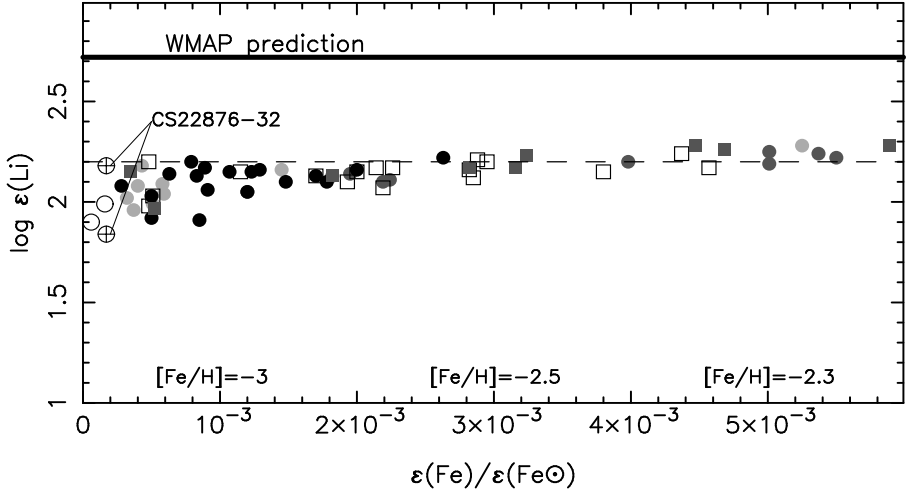}}
 \caption{$\rm log \epsilon(Li)$ vs. the iron abundance $\rm
 \epsilon (Fe_{*})/\epsilon(Fe_{\odot})$(linear scale) for turnoff stars  (\Teff $>5800$K).
 The symbols are the same as in Fig. 1.
 To improve the homogeneity, we have kept only the measurements of the lithium
 abundance based on a temperature determination independent of the
 reddening ($\rm T_{Hydrogen}$ or $\rm T_{exc}$). 
  When  $\rm \epsilon(Fe_{*})/\epsilon(Fe_{\odot}> 0.001$ ($\rm [Fe/H]>-3$), the scatter around the mean curve is very small (0.05 dex)  but at a lower metallicity the scatter suddenly increases.
   The lithium abundances in the
 two components of the extremely metal-poor binary CS~22876-32 are different by a factor of two.
 Lithium would sometimes suffers an extra depletion in the atmospheres of the most 
 metal-poor stars. Then, the pristine value of the lithium abundance would correspond to the higher envelope,   $\rm log \epsilon(Li) \approx 2.2$.}
\label{life}
\end{center}
\end{figure}

In Fig. \ref{Figlit} the scatter around the mean value of the lithium abundance is 0.10 dex. This scatter is not symmetric (see the histogram in Fig.  \ref{Figlit})  and is larger than the measurement errors. It is NOT mainly due to systematic differences between different authors, it exists even for the data of a given author.\\

What is the cause of this scatter of the lithium abundance when $\rm [Fe/H] < -2$ ?\\

If we plot  $\rm log \epsilon(Li)$ versus [Fe/H]  (Fig. 2)  a slight decrease of
$\rm \epsilon(Li)$ with the metallicity appears.  As a consequence the scatter observed in Fig \ref{Figlit} should be partly due to a dependence of lithium abundance on metallicity. 
This decrease suggests that the extrapolated abundance of lithium could be even lower than 2.15   when extrapolating toward zero-metal stars (just after the Big Bang).\\

In Fig. 3 we have plotted $\rm log \epsilon(Li)$
versus $\rm \epsilon (Fe)_{*} / \epsilon(Fe)_{\odot} $ (linear scale), 
and  for a better homogeneity, we have kept only stars
with a temperature  determined spectroscopically (H$\alpha$
profiles or excitation temperature) and thus independent of the interstellar
reddening.  
In the interval  $\rm 2~x~10^{-3} < \epsilon(Fe)_{*} / \epsilon(Fe)_{\odot} < 6~x~10^{-3}$ the lithium abundance increases steadily and the scatter around the mean curve is extremely small: 0.05 dex.

In the most metal-poor stars  ($\rm \epsilon(Fe)_{*} / \epsilon(Fe)_{\odot} < 2~x~10^{-3}$, i. e. $\rm [Fe/H]< -2.7$,  the mean value of the lithium abundance decreases suddenly to reach  $\rm log \epsilon(Li) \approx 2.0$  dex at $\rm \epsilon(Fe) = 0.$ (zero-metal stars). Therefore  $\rm log \epsilon(Li) \approx 2.0$ could be the mean value of the lithium abundance in the early Galaxy.
But in this region the scatter at a given metallicity increases strongly  (see  Sbordone et al. \cite{SBC10a} and  \cite{SBC10b} for confirmation by additional data).
This scatter could be interpreted as the result of a variable depletion, decreasing the lithium abundance from a natal value of about  $\rm log \epsilon(Li) = 2.2$ dex, this value  would represent the pristine value of the lithium abundance in the early Galaxy.\\
In both cases the pristine value of the lithium abundance is well below the cosmological value derived from the standard Big Bang (with the WMAP value of the $\eta$ parameter). 

In Fig 3, no stars are found above  $\rm log \epsilon(Li) = 2.2$ dex for $\rm [Fe/H]< -2.7$ and above  $\rm log \epsilon(Li) = 2.3$ dex for $\rm -2.7<[Fe/H]< -2.2$ defining an upper envelope of the lithium abundance. (A handful of Li-rich halo stars have been found in the literature, but all are several times more metal-rich than the stars considered here.)

Several interpretations can be put forward for explaining the  gap  between the cosmological and the observed lithium and also the sudden scatter appearing at low metallicity.\\

%ASTRATION
\subsection{Astration}  \label{Astration}
It has been proposed that the "uniform"   lithium abundance in halo dwarfs could be due to a general process affecting the whole Milky Way lowering the cosmological Li abundance down to the pristine abundance. Piau et al. (\cite{PBB06}) have proposed an early astration in very masssive stars Pop III stars (destroying Li).
They could have destroyed  3/4  of the primordial lithium, 50\% of the mass of the Galactic halo having been processed in these massive stars.  

The scatter of the lithium abundance appearing in the most metal-poor stars could then be explained by a still incomplete mixing of the matter in the Galaxy, or by the fact that these most metal-poor  stars would
have been accreted from different faint dwarf galaxies or satellites (Frebel et al., \cite{FSG10}) with different astration histories.
There are objections to a strong initial  astration in our Galaxy:  the ejecta of the stars destroying Li would have induced (Prantzos, \cite{Pra07}) a large early abundance of elements (e. g. C and O) that is  not observed. 
But, in $\rm \omega~Cen$, remnant of an accreted dwarf galaxy, the lithium abundance 
($\rm log \epsilon(Li) \approx 2.19$, Bonifacio et al., \cite{BMSs268}), is very similar 
to the Galactic value for stars of the same metallicity, although they certainly had different astration histories.
Probably, a very early global astration  of the matter of the universe, by massive first stars, before the formation of the galaxies,  would explain the plateau, but would encounter a similar objection about the early abundances of C and O in the Galaxy. However il could be possible that these massive Pop III stars rotate and thus eject only external layers with hydrogen and helium   (Meynet , this symposium).

%DEPLETION
\subsection{Depletion in the stellar atmospheres}  \label {Depletion}
 The gap between the theoretical value of the lithium abundance predicted by the standard Big Bang and the observed value in the atmosphere of the stars, can be also the result of a depletion of lithium, even in the warm metal-poor dwarfs.

The stellar abundance now observed could be due to diffusion (Michaud et al., \cite{MFG84}). Some calculations have been made (e. g. Richard et al., \cite{RMR05}): a  depletion such  as figured  in  Fig, 3, could be matched by  a  diffusion carefully moderated by some turbulence. The very low scatter of the data around the mean (0.05 dex) in the interval     $\rm 2~x~10^{-3} < \epsilon(Fe)_{*} / \epsilon(Fe)_{\odot} < 6~x~10^{-3}$ 
brings severe constaints to the theoretical computations. Also the sudden downward scatter, at the left part of the Figure 3  needs to be  explained.

The computation of lithium depletion  by diffusion in a globular cluster (Korn et al, \cite{KGR07}) carefully combined with turbulence, shows that the lithium depletion  amounts to about 0.26 dex. If this correction is applied to the most metal-poor field stars the pristine (or natal) Li abundance of these stars should be (in the best case) about 2.46 dex, a value still significantly smaller than the cosmological  value. As a consequence the depletion by diffusion moderated by turbulence cannot explain completely the gap. Moreover this depletion by diffusion/turbulence in globular clusters is questionned by Gonz\'alez Hern\'andez et al. (\cite{GBC10}), in this symposium.

We have to  remark (Fig. \ref{lilogfe} and \ref{life})  that  the two components of the extremely metal-poor ($\rm
[Fe/H]\approx-3.8$) binary star CS~22176-32, analyzed by Gonz\'alez Hern\'andez et
al. (\cite {GBL08}) have a lithium abundance different by a factor of
two, although their temperature is higher than 5800 K and that, as a consequence, no depletion of lithium is expected in these stars. 
This difference is above the measurement errors. The two stars were presumably born
with the same lithium abundance and thus the only explanation of this
large difference is that lithium has been depleted (at least) in CS~22176-32B.

"Extra-scatter" by "extra depletion'' in the  extremely metal-poor stars? This "extra-depletion''  would appear only, and sometimes, at very low  metallicity ($\rm [Fe/H]<-3$).

%\FIG 4
\begin{figure}[ht]
% \vspace*{-2.0 cm}
\begin{center}
\resizebox{8.0cm}{5.0cm}
{\includegraphics{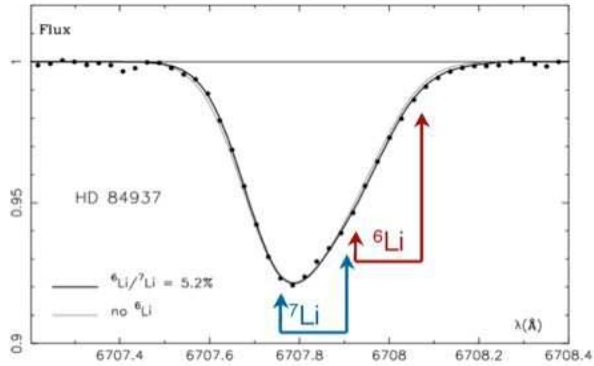}}
%{\includegraphics{fi-li6.eps}}
% \vspace*{-1.0 cm}
 \caption{Profile of the doublets of \7li and \6li in a very metal 
 poor turnoff star HD 84937, compared to a synthetic spectrum computed 
 with no \6li and \6li/7li=5\%
 }
\label{li6prof}
\end{center}
\end{figure}

%%%% SECTION 4
\section{Observation of \6li in the most metal poor stars in the Milky Way }-
Since \6li is not significantly formed by the standard Big Bang, if
\6li is observed in very metal-poor stars it is supposed to be formed
mainly by cosmic rays.

If \6li exists in these stars, its abundance is very small and it is thus very difficult to detect, in particular because the lines (doublets) of \6li and \7li are overlapping (see Fig. \ref{li6prof}). Very high resolution and high S/N  spectra are required as also very precise computations of the profiles. 

Several groups have recently tried to measure \6li in
metal-poor halo stars (see in particular Asplund et al. (\cite{ALN06}),
Asplund \& Mel\'endez (\cite{AM08}), Cayrel et al. (\cite{CSC07}), Garc\'{i}a P\'erez et al. (\cite{GAI09})).

Asplund et al. (\cite{ALN06}, \cite{AM08})   observed a sample of 27 turnoff stars with 
$\rm -3.3<[Fe/H]<-1.2$. In twelve of them,
they detected \6li. The abundance of \6li was constant 
$\rm log \epsilon(^6Li) \approx 0.8$ dex and the ratio \6li/\7li was between 4 and 10\%.
In all the 27 stars the abundance of lithium or its upper limit is compatible with a plateau with 
 \6li/\7li $ \approx $  5\%.  
  
This result remains fragile:  Cayrel et al. (\cite{CSC07}) have shown that, 
if a classical analysis of a given star would provide  a ratio \6li /\7li= 4\% the result would be 
 \6li /\7li= 0\% if  the asymetry of the lines, due to convection, is taken into account 
(3D computations).  

Recently, Garc\'{i}a P\'erez et al. (\cite{GAI09}) have insisted on the fact that the error of the
\6li measurement is generally underestimated due to the uncertainty of the position of the
continuum, the residual fringes etc.
It is interesting to remark that Asplund \& Mel\'endez (\cite{AM08}) found that the  \6li /\7li
ratio in G64-37 is 11 \% while Garc\'{i}a P\'erez et al. (\cite{GAI09}) found less than 1\%.

\subsection{Have we really observed \6li in EMP stars ? } 

In none of the stars, the \6li isotope is detected with a precision higher or equal to  $ 3 \sigma$ (Steffen et al., this symposium).  
If the error is a little underestimated, and if there is in fact, no \6li in the old metal-poor stars,  the noise would mimic an absorption of \6li  in about half of the stars  (detection) and an emission of \6li in the other half of the stars (no detection). 
This is not far from what is observed...

More precise computations (NLTE, 3D) about spectra with higher S/N ratios and a better defined continuum are necessary to firmly conclude about the detection and the abundance of \6li in the most metal-poor stars.

%%%%%Section 5 CONCLUSION
\section{Conclusion : Is it possible to explain the behaviour of \6li and \7li in the early Galaxy ?}

\subsection{\6li}
The absence of \6li, or at least an abundance below the limit of detection, would be easy to explain since the quantity of \6li formed by the standard Big bang and by the cosmic rays is supposed to be very low. 

The measurement of the \6li abundance is difficult. If some \6li is observed, it should have been formed (before the birth of our old turnoff stars)   by Galactic Cosmic Rays or during the explosion of massive supernovae in superbubles: we  should then observe a clear increase of  \6li with [Fe/H] like the increase of boron and beryllium (see Boesgaard, \cite{Boe04}, Prantzos, \cite{Pra07}).
At the present time, a large abundance of \6li, more or less constant with [Fe/H], such as observed by Asplund et al. (\cite{ALN06}, \cite{AM08}):  $\rm log \epsilon(Li) = 0.8 dex$, cannot be understood in the frame of the standard theories. 

\subsection{\7li}
Depending on the interpretation of Fig. \ref{life}, the pristine stellar abundance of \7li in our Galaxy is 
$\rm log \epsilon(Li)$ =  2.0 or 2.2 dex: in both cases,  it is far from the value (2.72 dex) of the \7li abundance produced by the  standard Big Bang nucleosynthesis, according  to the WMAP specifications. 
Some explanations can be proposed none of them completely satisfactory.
   
$\bullet$ {\bf Early astration} - 
According to   Piau et al. (\cite{PBB06}),  the primordial  \7li, (such as computed  by the Big Bang) would have been severely destroyed in the Galaxy before the formation of the old metal-poor stars by astration  of the matter in massive Pop~III stars, lowering the lithium abundance to the observed pristine abundance. But this theory cannot explain that  the same pristine abundance of lithium is observed in the Milky Way and in $\rm \omega Cen$ (the remnant of an accreted dwarf galaxy) which certainly had different astration histories  (see section \ref{Astration}).
Moreover in this case a strong excess (not observed) of C and O in the matter of the early Galaxy would be, a priori, expected,
However it has been remarked (Meynet , this symposium)  that massive rotating Pop~III stars would eject only external layers with hydrogen and helium (without excess of C and O ).

$\bullet$  {\bf Depletion / Diffusion} -
A second hypothesis is that lithium has been depleted in the atmosphere of the observed turnoff stars during their long evolution, by atomic diffusion carefully partially compensated by turbulence  (see section \ref{Depletion}).
%Richard, Michaud and Richer  (\cite{RMR05}) have proposed an  explanation of the plateau by  a diffusion carefully (partially) compensated by turbulence, reaching an approximate fitting of the computations of Li abundances to the  observations of the plateau.
 The data gathered in Fig. 3, which show for $\rm [Fe/H]>-3$ a very small scatter (entirely explained by the determination errors) are a challenge to the computations of such an uniform depletion. The downward scatter of the lithium abundance in the extremely metal-poor stars remains to be explained (variable extra-depletion ?), as well as the absence (within our limits of \Teff and metallicity) of any (Li-rich) star in the "desert" between the high Big Bang lithium abundance and the plateau. Anyway, if  the ``depletion correction'' computed by Korn et al. (\cite{KGR07}) for the globular cluster NGC 6397 is applied to the field metal-poor dwarfs, the resulting value of the pristine lithium abundance (in the best case, about 2.46 dex) is again far from the cosmological predictions.

$\bullet$  {\bf Gravity waves} -
The behaviour of lithium should be evaluated taking into account  the gravity waves (e. g. Charbonnel \& Talon \cite{TC04AA}, see also Talon et al., this symposium):  the results of  this promising theory  are eagerly awaited.

$\bullet$  {\bf Uncertainties in the Big Bang theories} -
The production of \7li  in the Big Bang 
could be lower in the standard nucleosynthesis  if a resonance level (Cyburt \& Pospelov, \cite{CP09}) is taken ito account in the transformation of  $^7$Be  into $^9$B. \\
Moreover, a number of theories have been proposed for various more or less  {\it ad hoc} non-standard Big Bang nucleosyntheses (see e. g. Iocco et al. \cite{IMM09}).\\

As a conclusion, in addition to progresses in fundamental physics and nucleosynthesis, the use of NLTE, 3D model atmospheres for interpreting high quality  spectra, analysing differential observations (comparison dwarfs-subgiants, binaries, delicate trends, small scatters and other such details) should shed some light on the always complex problem of the lithium abundance in the early Galaxy.

Acknowledgments
This text benefitted from conversations with P. Bonifacio, L. Sbordone and E. Caffau 

%%%%%%%%%%%%%%%%%%%%%%%%%%%%%%%%%%%

\end{document}